\documentstyle[aps,prl,psfig,floats]{revtex}
\begin{document}

\draft \voffset 1cm

\twocolumn[\hsize\textwidth\columnwidth\hsize\csname %
@twocolumnfalse\endcsname

\title{Evidence for Surface Andreev Bound states in Cuprate Superconductors from Penetration Depth Measurements}
\author{A. Carrington$^{1,2}$, F. Manzano$^{1,2}$, R. Prozorov$^3$, R. W. Giannetta$^3$, N. Kameda$^4$ and T. Tamegai$^4$ }
\address{$^1$Department of Physics and Astronomy, University of Leeds, Leeds LS2 9JT, England}
\address{$^2$H.H. Wills Physics Laboratory, University of Bristol, Bristol BS8 1TL, England}
\address{$^3$Department of Physics, University of Illinois at Urbana-Champaign, 1110 West Green St., Urbana, 61801 Illinois}
\address{$^4$Department of Applied Physics, The University of Tokyo, Hongo, Bunkyo-ku, Tokyo 113-8656, Japan.}
\date{\today}
\maketitle
\begin{abstract}
Tunneling and theoretical studies have suggested that Andreev bound states form at certain surfaces of unconventional
superconductors. Through studies of the temperature and field dependence of the in-plane magnetic penetration depth
$\lambda_{ab}$ at low temperature, we have found strong evidence for the presence of these states in clean single
crystal YBCO and BSCCO. Crystals cut to expose a \protect{[110]} interface show a strong upturn in $\lambda_{ab}$ at
around 7~K, when the field is oriented so that the supercurrents flow around this surface. In YBCO this upturn is
completely suppressed by a field of $\sim 0.1$~T.
\end{abstract}
 ]


The low temperature behavior of the magnetic penetration depth ($\lambda$) has played an important role in identifying
the bulk order parameter in the high $T_c$ cuprates.  Within the usual BCS quasiparticle picture, the temperature and
field dependence of $\lambda$ are determined by the energy dependence of the quasiparticle density of states $N(E)$,
and hence give information about the magnitude of the energy gap, but not its phase. Recently however, it has been
suggested that there is another contribution to $\lambda$ which arises from current carrying, zero energy, surface
Andreev bound states (ABS) \cite{hu94}, which form if there is $\pi$ phase shift between the different lobes of the
order parameter \cite{fogelstrom97,walter98,barash00}. The observation of these states provides a key piece of evidence
backing the $d$-wave scenario in the cuprates.  In addition, this extra contribution has important consequences for the
interpretation of all surface impedance measurements in non-conventional superconductors.

The zero bias conductance peak (ZBCP) observed in tunneling measurements has been interpreted as resulting from ABS
\cite{covington97,fogelstrom97}.  Theoretically, ABS are expected to form most strongly on [110] surfaces, with the
amplitude reducing monotonically to zero as the surface rotates to the [100], [010] or [001] directions. Experimentally
however \cite{covington97}, the ZBCP is seen for tunneling into all surfaces except [001].  This has been interpreted
as resulting from surface roughness and/or nanofaceting \cite{fogelstrom97}.  The contribution of the ABS to $\lambda$
can be understood simply by noting the increase in the penetration depth $\Delta \lambda$ due to thermally excited
quasiparticles is given by
\begin{equation}
\frac{\Delta \lambda (T)}{\lambda(0)}= \int_{-\infty}^\infty \frac{N(E)}{N(0)} \frac{\partial f}{\partial E} dE
\label{eq1}
\end{equation}
where $f$ is the Fermi function.  For a $d$-wave superconductor $N(E)\sim |E|$, which gives
$\Delta\lambda(T)/\lambda(0)=\alpha T$, where $\alpha$ depends on the angular slope of the energy gap near the
nodes\cite{yip92}.  Surface bound states add a singular contribution $\delta(E)$ to $N(E)$ which when substituted into
Eq.\ (\ref{eq1}) adds a divergent $1/T$ term to $\Delta \lambda (T)$. The relative size of $1/T$ term depends on the
orientation of the surface and the band structure of the material \cite{barash00}.  The ABS contribution to $\lambda$
is highly non-linear in field, as the superflow Doppler shifts of the position of the bound state peaks.  It is
therefore expected that the $1/T$ term is suppressed with a relatively weak magnetic field.

In this letter, we report measurements of the temperature and field dependence of the in-plane penetration depth of
high quality YBa$_2$Cu$_3$O$_{6.9}$ and Bi$_2$Sr$_2$CaCu$_2$O$_8$ single crystals which provide strong evidence of the
existence of ABS in these materials. Evidence for an ABS contribution to $\lambda$ have previously been identified from
studies of irradiated thin films \cite{walter98} and grain boundary junctions \cite{alff98}. Here, we have been able to
investigate the ABS more fully by measuring both the field and temperature dependence of $\lambda$ and by studying
clean single crystals which give us more control over the direction of the Meissner currents and are free from
intrinsic defects.

We have identified the ABS contribution to $\lambda$ in YBCO by three different techniques.  First, we have exploited
the anticipated strong dependence of the ABS term on the crystal surface orientation by measuring several crystals
which have been cut to expose different surfaces. Second, each crystal was measured in two orientations, with the probe
field $h$  parallel to either the $ab$-plane or the $c$-axis. Although the in-plane penetration depth is probed in both
these orientations, the Meissner currents flow along quite different surfaces in the two cases. For $h\|a$ or $b$ the
currents flow mostly along [001] surfaces where no ABS contribution is expected, whereas for $h\|c$ the currents flow
around the edges of the crystal and ABS contributions are expected provided that a non [100]/[010] surface is present.
This field orientation dependence also provides an important check for impurity effects or isotropic paramagnetic
contributions. Third, direct evidence for the presence of ABS states can be inferred from our measurements of the field
dependence of $\lambda_{ab}$.

Single crystals of YBCO were grown in yttria stabilized zirconia crucibles \cite{stupp92}, and were annealed for 3
weeks at 500$^\circ$C in flowing oxygen to obtain optimal doping ($T_c\simeq 94 $K, width $\sim 0.2$ K). Optical
microscopy and x-ray diffraction showed that the crystals selected for this work were $>$90\% untwinned.  The BSCCO
crystal was grown by the floating zone method and was annealed at 350$^\circ$C in 0.1\% 0$_2$ to produce a slightly
underdoped composition with $T_c=91$~K \cite{ooi98}. Measurements of changes in $\lambda$ as a function of temperature
and field were performed with resonant LC circuit driven by a tunnel diode operating at 11.7~MHz\cite{carrington99},
with frequency stability of a few parts in $10^{10}$~Hz$^{-\frac{1}{2}}$.  The dc field $H$ is always collinear with
the RF probe field $h$.

The YBCO samples were measured in three geometries with the field applied parallel to either the $a$, $b$ or $c$ axes.
For $h$ parallel to either $a$ or $b$ the measured frequency shift is directly proportional to $\Delta \lambda_{a,b}$
\cite{lamcnote}. In the third configuration ($h\|c$) the calibration is more difficult because of the large
demagnetizing factor. The calibration factor can be found either by normalizing the slope of the data to that for
$h\|a,b$ for 20~K$<T<$30~K where we expect the two configurations to be measuring the same quasiparticle response, or
by using a recently developed approximate analytic solution for the screening currents in this geometry
\cite{prozorov00}.  We find that both methods agree to within experimental error ($\pm 5$\%). We denote the change in
penetration depth relative to its value at our base temperature by $\Delta \lambda_{ab}^{a,b}(T)$ and $\Delta
\lambda_{ab}^c(T)$ for $h\|{a,b}$ and $h\|c$ respectively.

\begin{figure}
\centerline{\psfig{figure=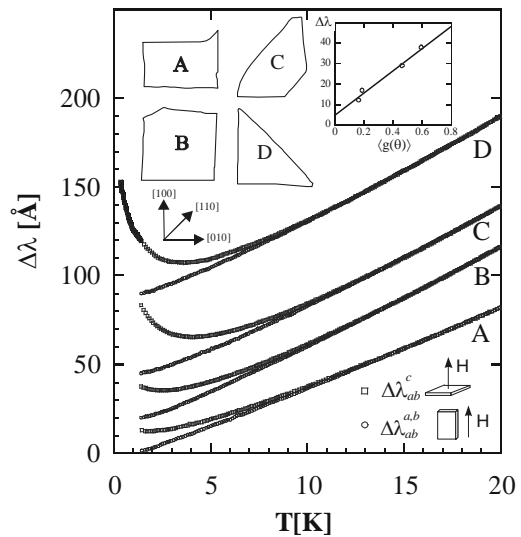,width=7cm}} \vskip 0cm
 \caption{Temperature dependence of the in-plane
penetration depths $\Delta \lambda_{ab}^{a,b}(T)$ and $\Delta \lambda_{ab}^c(T)$ for four crystals of YBCO. A constant
has been added to $\Delta \lambda_{ab}^c(T)$ so that is overlaps with $\Delta \lambda_{ab}^{a,b}(T)$ at high
temperature. The data for each crystal have been offset for clarity. The shape of the crystals is indicated in the
figure. For sample D data taken down to 0.35~K in another apparatus is shown.  The inset shows the difference between
$\Delta \lambda_{ab}^{a,b}(T)$ and $\Delta \lambda_{ab}^c(T)$ at 1.4~K versus the shape factor $\langle
g(\theta)\rangle$.} \label{ybcofig}
\end{figure}

The temperature dependence of $\lambda_{ab}$ for four samples of YBCO in the two field orientations is shown in Fig.\
\ref{ybcofig}. It can seen that for all samples $\Delta \lambda_{a,b}^{ab}(T)$ is highly linear at low temperature
indicating high sample quality \cite{labcnote,hirschfeld93}.  However, the data in the second field orientation
[$\Delta \lambda_{ab}^c(T)$] is markedly different below $\sim 7-10$K.  There is also a clear correlation between the
size of this upturn in $\Delta \lambda_{ab}^c(T)$ and the shape of the sample. In the sequence of samples from
A$\rightarrow$D the proportion of non [100]/[010] surface increases along with the size of upturn in $\lambda(T)$.
These strong orientation dependencies of the upturn are clear evidence for it originating from ABS.

Detailed calculations of the ABS contribution to $\lambda$, including the effects of ABS broadening, have been
performed by Barash {\it et al.} \cite{barash00} For a simple cylindrical Fermi surface the magnitude ($\beta$) of the
ABS $1/T$ term was calculated to be $\hbar v_f g(\theta)/[6k_B\lambda(0)]$, where
$g(\theta)=||\cos^3(\theta)|-|\sin^3(\theta)||$ is the dependence of the upturn on the angle ($\theta$) the surface
makes with the [110] direction, and $v_f$ is the Fermi velocity.  The effect of a finite ABS lifetime is to broaden the
$\delta$-function contribution to $N(E)$ making the divergence of $\lambda(T)$ weaker. The results of the numerical
calculations in Ref.\ \cite{barash00} can be approximated (for $T \gg 0$) by replacing the $1/T$ term by $1/(T+T^*)$.

By subtracting the data for $\Delta \lambda_{ab}^{a,b}(T)$ from $\Delta \lambda_{ab}^c(T)$ we can isolate the ABS
contribution. Fitting this to $c/(T+T^*)$ gives $T^*=0.8\pm$0.4K and $c=100\pm30$ \AA K for samples C and D, implying
that the ABS states are only slightly broadened. The calculation of Barash {\it et al.} yields a somewhat larger value
$c=10^3$~\AA K for these crystals (assuming a cylindrical FS and $v_f=2\times 10^5$~ms$^{-1}$), but we note that many
factors, such as band-structure or diffuse scattering, can reduce $c$ from this simple theoretical estimate.

We have attempted to test the correlation of the size of the upturn in $\lambda$ with the crystal shape in a more
quantitative manner by calculating $\langle g(\theta)\rangle$ for our crystals, where $\langle\rangle$ denotes an
average around the surface of the crystal. $\theta$ was calculated from a digitized optical image of the crystal. The
inset to Fig.\ \ref{ybcofig} shows $\Delta \lambda_{ab}^c(T)-\Delta \lambda_{ab}^{a,b}(T)$ at 1.4~K as a function of
$\langle g(\theta)\rangle$.  The linear fit shows that the correlation is good and the small intercept at $\langle
g(\theta)\rangle$=0 indicates that there is not a significant effect of microscopic surface roughness below our optical
resolution.

\begin{figure}
\centerline{\psfig{figure=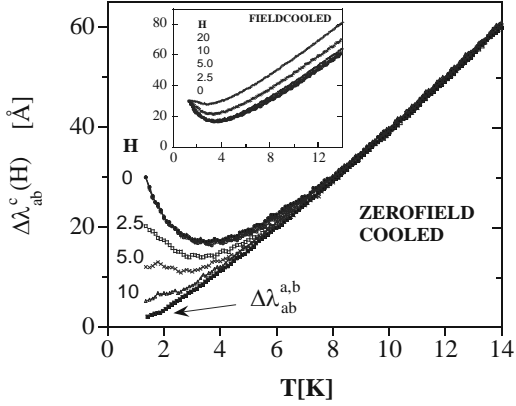,width=7cm}} \caption{ZFC data for $\Delta \lambda_{ab}^c(T,H)$ for sample D in
fixed fields between 0 and 10 mT, along with $\Delta \lambda_{ab}^{a,b}(T)$ in zero field.  The inset shows FC data for
the same crystal.} \label{tsweepbfig}
\end{figure}

Next we turn to the field dependence of $\lambda$.  We have investigated this in two ways; either sweeping the
temperature at fixed field or sweeping the field at fixed temperature.  The results of the temperature sweeps at fixed
field are shown in Fig.\ \ref{tsweepbfig}.  A constant has been added to the $\Delta\lambda_{ab}^c(T)$ data in finite
field so that they overlap with the zero field data at high temperature.  By plotting the data in this way it can
clearly be seen that an applied field suppresses the low temperature upturn in $\lambda(T)$. The good collapse of the
data for $T>10$~K implies that the field dependence of $\lambda$ [$\Delta \lambda(H)$], above 10~K is temperature
independent. For each run the sample was cooled in zero field (ZFC) and then the field was applied at 1.4~K.  The
temperature was then cycled up to 30~K several times to check for drift and any hysteresis.  For fields $\leq$ 10 mT no
hysteresis was seen and a zero field run following the finite field sweep overlap the virgin zero field curve exactly.
We therefore conclude that for $\mu_0 H\leq 10$~mT no vortices leak into the sample and we are in a reversible Meissner
state. It can be seen that a field of $\sim 10$ mT is sufficient to suppress the upturn. For fields higher than $\sim$
15 mT hysteresis was observed in the first $T$ sweep after which the data became reversible. This hysteresis is
presumably caused by flux initially leaking into the sample but becoming pinned away from the edges after the
temperature cycle. Data were also taken where the sample was cooled through $T_c$ to low temperature with the field
applied (denoted FC, inset Fig.\ \ref{tsweepbfig}). The effect of field in this case was much reduced (by approximately
a factor of 4). This is expected because of the more homogenous flux profile and thus reduced surface field in this
case.

The effect of a dc magnetic field on the ABS contribution can be understood as originating from a Doppler shift $dE$ in
the quasiparticle energy spectrum, where $dE=e\vec{v}_f\cdot\vec{A}$ ($\vec{A}$ is the vector potential).  The effect
of this shift on the ABS peaks has been observed directly in tunneling measurements
\cite{fogelstrom97,covington97,aprili99}.  Substituting $N(E)=\delta(E + e\vec{v}_f\cdot\vec{A})$ into Eq.\ (\ref{eq1})
gives
\begin{equation}
\frac{\Delta \lambda(T,H)}{\lambda(0)} = \frac{\beta}{4T} \cosh^{-2}\left[\frac{\mu_0 e \lambda \vec{H}\cdot\vec{v_f}
}{2k_BT}\right]
 \label{eq2}
\end{equation}
showing that the $1/T$ divergence of $\lambda$ is suppressed when the applied field exceeds the temperature dependent
field scale $\widetilde{H}=k_BT/(\mu_0ev_f\lambda)=H_0T/T_c$, where $H_0$ is of order the thermodynamic critical field
$\mu_0 H_c=\Phi_0/\lambda \xi$.  A more exact formula for $\Delta\lambda(H)$, which includes the necessary averaging
over all quasiparticle trajectories has been given in Ref.\ \cite{barash00}. Numerically, the result does not differ
significantly from Eq.\ (\ref{eq2}) except in the effective value of $\widetilde{H}$. There is also a contribution from
the shift of the non-singular part of the quasiparticle spectrum but this is considerably weaker [$\Delta
\lambda/\lambda(0) \sim H/H_0$] \cite{yip92}.

\begin{figure}[t]
\centerline{\psfig{figure=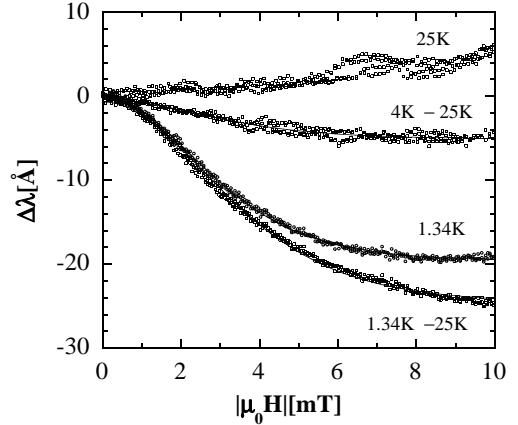,width=7cm}} \vskip 0cm \caption{ ZFC Field sweeps at fixed temperature for sample
D. For two of the curves the 25~K data have been subtracted.  The solid lines are fits to Eq.\ (\ref{eq2}).}
\label{bsweeptfig}
\end{figure}

To make a detailed comparison to the data we first note that as the sample has a large aspect ratio, the effective
surface field will be significantly enhanced above the applied field.  Also, we expect that the field will also vary
somewhat across the thickness of the crystal, achieving its maximum value near the top faces. We estimate an average
{\lq}demagnetizing{\rq} factor $\eta_{av}=1/(1-N_{av})$ by comparing the total frequency shifts when we extract the
sample out of the coil in the $H\|a,b$ and $H\|c$ geometries. For sample D, $\eta_{av}=11.7$.

The results of field sweeps at fixed temperature are shown in Fig.\ \ref{bsweeptfig}. At 1.34~K a very strong decrease
in $\lambda$ with field is observed.  Above ~10K, $\Delta\lambda(H)$ becomes positive and, to a good approximation,
temperature independent.  For samples which do not show a strong upturn in $\Delta \lambda_{ab}^c(T)$ (e.g., sample A)
this behavior persists down to 1.3~K \cite{carrington99}.  We conclude therefore, that this behavior is a response of
the bulk supercurrents.  The response bears some resemblance to the prediction of Yip and Sauls for the non-linear
Meissner effect \cite{yip92}. However, the fact that it is largely temperature independent is in serious disagreement
with the theory. This effect is nevertheless small in comparison to the total response at low temperature for samples C
and D. The lower curve in Fig.\ \ref{bsweeptfig} shows $\Delta\lambda$(H,T=1.35~K) with the temperature independent
contribution subtracted along with a fit to Eq.\ (\ref{eq2}). It can be seen that the agreement is excellent, with
$\mu_0\widetilde{H}=1.9$~mT.  When allowance is made for the field enhancement $\mu_0\widetilde{H}_{eff}\simeq
22.2$~mT, implying  $v_f =0.4\times10^5$ms$^{-1}$.  A fit to the Barash formula \cite{barash00} gives a smaller value
of $\mu_0\widetilde{H}=0.86$~mT and a larger value of $v_f =0.86\times10^5$ms$^{-1}$.  At higher $T$ we find that
$\widetilde{H}$ increases but not as strongly as would be expected from Eq.\ (\ref{eq2}).  Qualitatively, this behavior
was predicted by Barash {\it et al.} who showed that $\widetilde{H}$ tends to a constant as $T\rightarrow T^*$.  Thus
at our lowest temperature $\widetilde{H}$ may be larger than it would be in the pure limit and the above analysis
underestimates the intrinsic value of $v_f$. With $T^*=0.8\pm0.4$~K the Barash formula gives $v_f
=1.2\pm0.2\times10^5$ms$^{-1}$.

\begin{figure}
\centerline{\psfig{figure=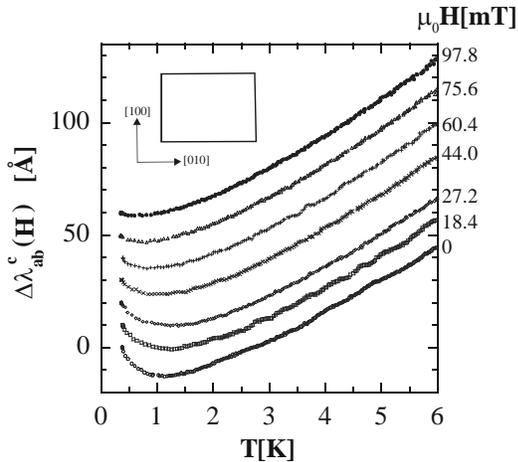,width=7cm}} \vskip 0cm \caption{Field cooled measurements of $\Delta
\lambda^c_{ab}(T)$ for a BSCCO single crystal. The data have been offset for clarity.  The inset shows the shape of the
sample.} \label{bsccofig}
\end{figure}

It is worth discussing briefly other processes which could effect the low temperature behavior of $\lambda$. Small
amounts of unitary scatters \cite{hirschfeld93}, and non-local effects \cite{kosztin97} can both cause $\lambda(T)\sim
T^2$ at low temperature, but do not cause an upturn. In some high $T_c$ cuprates a  $1/T$ term in $\lambda(T)$ has been
linked to the paramagnetic ions \cite{prozorov00a} or amorphous surface layers \cite{panagopoulos96}. It is highly
unlikely that such contributions would be present in clean single crystals of YBCO and in any case this effect would
not depend on the orientation of the exposed surface or the field orientation in the manner observed here. Also the
effect of field on such impurities would be much weaker than we have observed (e.g., $k_BT/g\mu_B\simeq1$~telsa, with
$g$=2 and $T$=1.4~K).

Measurements of BSCCO crystals show that similar effects occur in other cuprates. In Fig.\ \ref{bsccofig} we show
$\Delta\lambda_{ab}^c$ for a sample of BSCCO in fields up to 97.8~mT. In zero field an upturn is clearly visible below
$\sim$ 1~K and again the application of field suppresses this upturn. Because of the very strong anisotropy of BSCCO it
is not possible to measure $\lambda_{ab}(T)$ in the $H\|a,b$ configuration, although the absence of the upturn in this
geometry precludes paramagnetic impurities. The very weak pinning in BSCCO also precludes ZFC measurements, as here we
find irreversible behavior even in low fields. A larger field is required to suppress the upturn in this sample however
this is largely explained by the smaller effect of a field in the FC case. This upturn has only been seen in this very
clean sample of BSCCO. The smaller magnitude of the upturn [or the lower temperature of the minimum in $\lambda(T)$] is
explained by the surfaces of this crystal being close to [100]/[010].

In conclusion, by studying the effect of the orientation of the applied field to the crystal axes on the in-plane
penetration depth, we have shown strong evidence for the existence of current carrying Andreev bound states on non
[100] surfaces of YBCO.  In clean crystal the states are only weakly broadened.  The response is shown to be highly
non linear in the Meissner phase, with a small field completely suppressing the anomaly.  This behavior is consistent
with a shift of the ABS peak in the quasiparticle density of states caused by the screening currents.  Similar effects
have been observed in BSCCO indicating that these effects are universal in cuprates and possibly all non-conventional
superconductors.  Besides being of intrinsic interest we also note that these effects must be taken into account in
measurements of the surface properties of all non-conventional superconductors which have surfaces where Andreev bound
states can form.

We acknowledge useful discussions with M.\ Aprili, D.\ Lee, J.R.\ Cooper and L.H.\ Greene. Work at Urbana was supported
through State of Illinois - ICR funds.


\begin{references}
\bibitem{lamcnote} Contributions from $\lambda_c$ are negligible as the aspect ratios of our samples
are $>25$ \protect\cite{carrington99}.
\bibitem{labcnote} For the $H\|a,b$ configuration $\Delta \lambda_{ab}^{a,b}(T)$ is
calculated by averaging $\Delta \lambda_a(T)$ and $\Delta \lambda_b(T)$ according to the relative dimensions of the
crystal in the $a$ and $b$ directions.
\end{references}
\end{document}